\documentclass{pnastwo}

\usepackage{amssymb,amsfonts,amsmath}
\usepackage{rotating}
\usepackage{setspace}
\usepackage{lscape}
\usepackage{graphicx}

\contributor{Submitted to Proceedings
of the National Academy of Sciences of the United States of America}
\url{www.pnas.org/cgi/doi/10.1073/pnas.0709640104}
\copyrightyear{2010}
\issuedate{Issue Date}
\volume{Volume}
\issuenumber{Issue Number}

\begin{document}



\title{Dynamic reconfiguration of human brain networks during learning}





\author{Danielle S. Bassett\affil{1}{Complex Systems Group, Department of Physics, University of California, Santa Barbara, CA 93106, USA},
Nicholas F. Wymbs\affil{2}{Department of Psychology and UCSB Brain Imaging Center, University of California, Santa Barbara, CA 93106, USA},
Mason A. Porter\affil{3}{Oxford Centre for Industrial and Applied Mathematics, Mathematical Institute, University of Oxford, Oxford OX1 3LB, UK} \affil{4}{CABDyN Complexity Centre, University of Oxford, Oxford
    OX1 1HP, UK},
Peter J. Mucha\affil{5}{Carolina Center for Interdisciplinary Applied
    Mathematics, Department of Mathematics, University of
    North Carolina, Chapel Hill, NC 27599, USA}\affil{6}{Institute for Advanced Materials, Nanoscience \&
    Technology, University of North Carolina, Chapel Hill, NC
    27599, USA},
Jean M. Carlson\affil{1}{},
\and Scott T. Grafton\affil{2}{}}

\contributor{Submitted to Proceedings of the National Academy of Sciences
of the United States of America}

\maketitle

\begin{article}

\begin{abstract} Human learning is a complex phenomenon requiring flexibility to adapt existing brain function and precision in selecting new neurophysiological activities to drive desired behavior \cite{Buchel1999,Tunik2007}. These two attributes -- flexibility and selection -- must operate over multiple temporal scales as performance of a skill changes from being slow and challenging to being fast and automatic \cite{Newell2009,Doyon2005}. Such selective adaptability is naturally provided by modular structure \cite{Simon1962,Kirschner1998,Kashtan2005}, which plays a critical role in evolution \cite{Masel2010}, development \cite{Kirschner1998}, and optimal network function \cite{Kashtan2005,Bassett2010}. Using functional connectivity measurements of brain activity acquired from initial training through mastery of a simple motor skill, we investigate the role of modularity in human learning by identifying dynamic changes of modular organization spanning multiple temporal scales. Our results indicate that flexibility, which we measure by the allegiance of nodes to modules, in one experimental session predicts the relative amount of learning in a future session. We also develop a general statistical framework for the identification of modular architectures in evolving systems, which is broadly applicable to disciplines where network adaptability is crucial to the understanding of system performance. \end{abstract}






The brain is a complex system, composed of many interacting parts, which dynamically adapts to a continually changing environment over multiple temporal scales. Over relatively short temporal scales, rapid adaptation and continuous evolution of those interactions or \emph{connections} form the neurophysiological basis for behavioral adaptation or learning. At small spatial scales, stable neurophysiological signatures of learning have been best demonstrated in animal systems at the level of individual synapses between neurons \cite{Kim1997,Glanzman2008,Xu2009}. At a larger spatial scale, it is also well known that specific regional changes in brain activity and effective connectivity accompany many forms of learning in humans---including the acquisition of motor skills \cite{Buchel1999,Tunik2007}.

Learning-associated adaptability is thought to stem from the principle of cortical modularity \cite{Hart2010}. Modular, or nearly decomposable \cite{Simon1962}, structures are aggregates of small subsystems (modules) that can perform specific functions without perturbing the remainder of the system. Such structure provides a combination of compartmentalization and redundancy, which reduces the interdependence of components, enhances robustness, and facilitates behavioral adaptation \cite{Kirschner1998,Felix2008}. Modular organization also confers evolvability on a system by reducing constraints on change \cite{Kirschner1998,Kashtan2005,Wagner1996,Schlosser2004}. Indeed, a putative relationship between modularity and adaptability in the context of human neuroscience has recently been posited \cite{Meunier2010b,Werner2010}. To date, however, the existence of modularity in large-scale cortical connectivity during learning has not been tested directly.

Based on these aforementioned theoretical and empirical grounds, we hypothesized that the principle of modularity would characterize the fundamental organization of human brain functional connectivity during learning. More specifically, based on several studies relating the neural basis of modularity to the development of skilled movements \cite{Burdet1998,Sosnik2004,Schaal2005}, we expected that functional brain networks derived from acquisition of a simple motor skill would display modular structure over the variety of temporal scales associated with learning \cite{Doyon2005}. We also hypothesized that that modular structure would change dynamically during learning \cite{Buchel1999,Newell2009}, and that characteristics of such dynamics would be associated with learning success.

We tested these predictions using functional magnetic resonance imaging (fMRI), an indirect measure of local neuronal activity \cite{Lee2010}, in healthy adult subjects during the acquisition of a simple motor learning skill composed of visually cued finger sequences. We derived low-frequency (0.06--0.12 Hz) functional networks from the fMRI data by computing the temporal correlation between activity in each pair of brain regions to construct weighted graphs or whole-brain functional networks \cite{Bassett2010d,Bassett2009,Bassett2006} (See Supplementary Information for details). This network framework enabled us to estimate a mathematical representation of modular or community organization, known as `network modularity', for each individual over a range of temporal scales. We evaluated the evolution of network connectivity over time using a novel mathematical framework \cite{Mucha2010}, and we tested its relationship with learning using the Pearson correlation coefficient. See Methods for details of the sample, experimental paradigm, and methods of analysis.

\section{Results}
\subsection{Static Modular Structure}
We investigated network organization over multiple temporal scales---over days, hours, and minutes---during motor learning \cite{Newell2009,Doyon2005} (see Figure 1B), and we found it to be significantly modular at each scale (see Figure 2A-C). A diagnostic measure of the amount of network modularity in the system is the modularity index $Q$ (See Methods for a mathematical definition). Importantly, while an increase in the modularity index $Q$ indicates a significant segregation of system structure into distinct modules or communities, the number of modules that we uncovered is significantly smaller than that expected in a random system, indicating a complementary integration. The scaling of architectural organization over multiple temporal resolutions is consistent with previous findings for the human cortex in both the frequency \cite{Bassett2006a} and spatial domains \cite{Bassett2010,Bassett2010c}. Furthermore, the temporal structure of this architecture is graded in the sense that fewer modules (about $3$) on longer time scales (Figure 2A-B) are complemented by more modules (about $4$) on shorter time scales (Figure 2C). This graded structure is analogous to that found in the nested modular networks of underlying brain anatomy where few modules uncovered at large spatial scales are complemented by more modules at smaller spatial scales \cite{Bassett2010}.

\subsection{Dynamic Modular Structure}
We next consider evolvability, which is most readily detected when the organism is under stress \cite{Masel2010} or when acquiring new capacities such as during external training in our experiment. We found that the community organization of brain connectivity reconfigured adaptively over time. Using a recently-developed mathematical formalism to assess the presence of dynamic network reconfigurations \cite{Mucha2010}, we constructed multilayer networks in which we link the network for each time window (see Figure 3A) to the network in the time windows before and after (Figure 3B) by connecting each node to itself in the neighboring windows. We then measured modular organization \cite{Porter2009,Fortunato2010,Blondel2008} on this linked multilayered network in order to find long-lasting modules \cite{Mucha2010}.

To verify the reliability of our measurements of dynamic modular architecture, we introduced three null models based on permutation testing (Figure 3C). We found that cortical connectivity is specifically patterned, which we concluded by comparison to a `connectional' null model in which we scrambled links between nodes in each time window \cite{Maslov2002}. Furthermore, cortical regions maintain these individual connectivity signatures that define community organization, which we determined by comparison to a `nodal' null model in which we linked a node in one time window to a randomly chosen node in the previous and next time windows. Finally, we found that functional communities display a smooth temporal evolution, which we identified by comparing diagnostics derived from the true multilayer network structure to those derived from a temporally permuted version (see Figure 3D). We constructed this `temporal' null model by randomly re-ordering the multilayer network layers in time.

By comparing the structure of the cortical network to that of the null models, we found that the human brain displayed a heightened modular structure in which more modules of smaller size were discriminable as a consequence of the emergence and extinction of modules in cortical network evolution. The `stationarity' of communities, which is defined as the average correlation between partitions over consecutive time steps \cite{Palla2007}, was also higher in the human brain than in the connectional or nodal null models, indicating a smooth temporal evolution.

\subsection{Learning}
Given the dynamic architecture of brain connectivity, it is interesting to ask whether the specific architecture changes with learning---either at a gross scale through an adaptation in the number or sizes of modules or at a finer scale through alterations in the nodal composition of modules. Empirically, we found no significant differences between experimental sessions in the coarse diagnostics. To quantify finer-scale architectural fluctuations, we introduced the notion of node \emph{flexibility}, which we define using the multilayer framework to be the number of times that each node changes module allegiance, normalized by the total possible number of changes (see Supplementary Materials for details).  The flexibility of the network as a whole is then defined as the mean flexibility over all nodes.

Network flexibility is a biologically meaningful measure that captures changes in the local properties of individual network elements. We found that network flexibility changed during the learning process -- first increasing and then decreasing (see Figure 4A). In particular, the flexibility of a participant in one session could be used as a predictor of the amount of learning (as measured by the improvement in the time required to complete the sequence of motor responses) in the following session (Figure 4B). Regions of the brain that were most responsible for this predictive power of individual differences in learning were distributed throughout the cortex, with strong loadings in the medial, inferior, and pre-frontal, presupplementary motor, posterior parietal, and occipital cortices (Figure 4C-D). The ability to predict future learning capacity could not be reliably derived from fMRI activation, supporting our conclusion that flexibility provides a useful approach for modeling system evolvability.

Our results indicate that flexibility is sensitive to both intra-individual and inter-individual variability. Across participant, we found that network flexibility was modulated by learning (Figure 4A). However, we also found that each participant displayed a characteristic flexibility. The variation in flexibility over participants was larger than the variation in flexibility across sessions, as measured by the intra-class correlation coefficient ($ICC \approx 0.56$; F-statistic $F(17,34) \approx 4.85$, $p \approx 4 \times 10^{-5}$).

\section{Discussion}
\subsection{Modularity of Functional Connectivity}
Modularity is an intuitively important property for dynamic, adaptable systems. The accompanying system decomposability provides the necessary structure for complex reconfigurations. Modularity can be a property of morphology, as has been widely described in the context of evolution and development \cite{Masel2010,Wagner1996,Schlosser2004}, as well as of the interconnection patterns of social, biological, and technological systems \cite{Porter2009,Fortunato2010}. More pertinent to this paper, recent evidence suggests that modular organization over several spatial scales, or \textit{hierarchical modularity}, also characterizes the large-scale anatomical connectivity of the human brain \cite{Bassett2010,Bassett2010c}, as well as the spontaneous fluctuations \cite{Meunier2009,Meunier2010} thought to stem from anatomical patterns \cite{Damoiseaux2009}. However, the putative relationship between adaptability and modular structure has not been previously explored in the context of the brain connectome.

In the present study, we have shown that the functional connectivity of the human brain during a simple learning paradigm is non-homogenous. Rather, it is segregated into communities that can each perform unique functions. This segregation of connectivity structure was expressed consistently over the scale of days, hours, and minutes, suggesting that community structure provides a generalizable framework for the successful actuation of temporally distinct phenomena \cite{Schlosser2004}. However, it is also notable that connectivity at the shortest temporal scale displayed higher variability, perhaps reflecting the necessity for dynamic modulation of human brain function over relatively short intervals during learning \cite{Newell2009}. In light of historically strict definitions of cognitive modules as completely encapsulated structures \cite{Fodor1983}, it is important to emphasize that the modules that we have uncovered here remain integrated with one another by a complex pattern of weak interconnections.

\subsection{Dynamic Network Evolution}
Efforts to characterize both resting state \cite{Raichle2007} and task-based large-scale connectivity of human brain structure and function \cite{Bassett2010d,Bassett2009,Bassett2006} have focused almost exclusively on static representations of underlying connectivity patterns. However, both scientific intuition and recent evidence suggest that connectivity can be modulated both spontaneously \cite{Raichle2010} and by exogenous stimulation \cite{Buchel1999}. The exploration of temporally evolving network architecture therefore forms a critical new frontier in neuroscience.

Our exploration of dynamic community structure in an experimental paradigm that requires neurophysiological adaptability provides insight into the organizational principles supporting successful brain dynamics. Similar to social systems \cite{Palla2007}, we found that community organization changed smoothly with time, displaying coherent temporal dependence on what had gone before and what came after, a characteristic compatible with complex long-memory dynamical systems \cite{Achard2008}.

In addition to global adaptability, we found that diverse regions of the brain performed different roles within those communities: Some maintain community allegiance throughout the experiment (low flexibility nodes), and others constantly shift allegiance (high flexibility nodes). Biologically, this network flexibility might be driven by physiological processes that facilitate the participation of cortical regions in multiple functional communities. Learning a motor skill induces changes in both the structure and connectivity of the cortex\cite{Hofer2010,Scholz2009}, which is accompanied by increased excitability and decreased inhibition of neural circuitry \cite{Smyth2010,Ljubisavljevic2006,Beers2009}. However, it is plausible that flexibility might also be driven by task-dependent processes that require the capacity to balance learning across subtasks. For example, the particular experiment utilized in this study demanded that subjects master the performance of using the response box, decoding the stimulus, performing precise movements, shifting attention between stimuli, and switching between different movement patterns.

\subsection{Flexibility and Learning} Importantly, the inherent temporal variability in network structure measured by nodal flexibility was not a stable signature of an individual's functional organization but was instead modulated by consecutive stages of learning -- first increasing and then decreasing as movement time stabilized in the later stages of learning \cite{Newell2009}. The modulation of flexibility by learning was evident not only at the group level but also in individuals. The amount of flexibility in each participant could be used to predict that participant's learning in a following experimental session. In addition to supporting the theoretical utility of accessible but often ignored higher-order (bivariate, multivariate) statistics of brain function, this result could potentially be used to inform decisions on how and when to train individuals on new tasks depending on the current flexibility of their brain. From this work alone, however, we are unable to determine whether or not learning is the only possible modulator of flexibility. Complementary experiments could be designed to test whether flexibility is also modulated by fatigue or exogenous stimulants in order to increase subsequent skilled learning. We also found that inter-individual variability in flexibility was larger than intra-individual variability, indicating that flexibility might be a reliable indicator of a given subject's brain state. Consequently, our methodology could potentially be of use in predicting a given individual's response to training or neurorehabilitation \cite{Krakauer2006,Mulder2001}.

Flexibility might be a network signature of a complex underlying cortical system characterized by noise \cite{Faisal2008}. Such a hypothesis is bolstered by recent complementary evidence suggesting that variability in brain signals also supports mental effort in a variety of cognitive operations \cite{McIntosh2008}, presumably by aiding the brain in switching between different network configurations as it masters a new task. Indeed, the theoretical utility of noise in a nonlinear dynamical system like the brain \cite{Freeman1994} lies in its facilitation of transitions between network states or system functions \cite{Deco2009} and therefore helps to delineate the system's dynamic repertoire \cite{Lippe2009}. However, despite the apparent plausibility that network flexibility and cortical noise are related, future studies are necessary to directly test this hypothesis.

\subsection{Methodological Considerations}
The construction of brain networks from continuous association matrices, such as those based on pairwise correlation or coherence, has historically been performed by applying a threshold to the data to construct a binary graph in which an edge exists if the association between the nodes it connects was above the threshold and does not exist otherwise \cite{Bassett2010d,Bassett2009,Bassett2006}. However, the statistical validity of that method is hampered by the need to choose an arbitrary threshold as well as by the discretization of inherently continuous edge weights. In the current work, we have instead used fully weighted networks in which connections retain their original association value unless that value was found to be insignificant (based on statistical testing employing a false discovery rate (FDR) correction for multiple comparisons) \cite{Genovese2002}. Future studies comparing multiple network construction techniques will be important to statistically assess the added value of weighted edge retention in the assessment of network correlates of cognition.

Second, partitioning a set of nodes into a set of communities is NP-hard \cite{Brandes2008}, meaning that modularity algorithms produce many near-optimal partitions of the network \cite{Good2010}. The number of near-optimal partitions is higher in large networks, or those with edges weighted with only 1's and 0's \cite{Good2010}. Here we have chosen to study small weighted networks in which the number of near-optimal partitions is small. Nevertheless, we have systematically explored the partition landscape by iterative optimization. Accordingly, we report mean modularity estimates that our results suggest are representative of the system (see Supplementary Materials for further details). However, further work is necessary to measure common community assignments within the ensemble of partitions in order to identify consistently segregated groups of brain regions. This will aid in further exploration of the biological relevance of the uncovered communities.

Finally, the statistical validation of community structure in social and biological systems is complicated by several factors. For example, many investigations, especially in social systems, are hindered by their small number of instantiations. In our work, the relatively large number of subjects in conjunction with estimations of multiple networks over various temporal scales facilitated a stringent statistical assessment of community structure both in comparison to randomly connected graphs and, as we have developed for dynamic networks, to graphs where nodal identities or times were scrambled. An important future area of research will focus on the development of alternative null models that are not perfectly random but which assume increasingly biologically realistic network architectures.

\section{Conclusion}
Consistent with our hypotheses, we have identified significant modular structure in human brain function during learning over a range of temporal scales: over days, hours, and minutes. Modular organization over short temporal scales changed smoothly, suggesting system adaptability. The composition of functional modules displayed temporal flexibility that was modulated by early learning, varied over individuals, and was a significant predictor of learning in subsequent experimental sessions. Furthermore, we developed and reported a general framework for statistical validation of dynamic modular architectures in arbitrary systems. More generally, our evidence for adaptive modular organization in global brain activity during learning provides critical insight regarding the dependence of system performance on underlying architecture.



\begin{materials}
Twenty-five right-handed participants (16 female, 9 male; mean age 24.25 years) volunteered with informed consent in accordance with the UCSB Internal Review Board.  After exclusions for task accuracy, incomplete scans, and abnormal Magnetic Resonance Imaging (MRI), 18 participants were retained for subsequent analysis. All participants had less than 4 years of experience with any one musical instrument, had normal vision, and had no history of neurological disease or psychiatric disorders. Participants were paid for their participation.

The experimental framework consisted of a simple motor learning task in which subjects responded to a visually cued sequence by generating responses using the 4 fingers of their non-dominant hand (thumb excluded) on a custom response box. Participants were instructed to respond swiftly and accurately. Visual cues were presented as a series of musical notes on a 4-line music staff such that the top line of the staff mapped to the leftmost key depressed with the pinkie finger. Each 12-note sequence contained 3 notes per line, which were randomly ordered without repetition and free of regularities such as trills and runs. The number and order of sequence trials was identical for all participants. All participants completed 3 training sessions in a 5-day period, and each session was performed inside the MRI scanner.

Functional MRI (fMRI) recordings were conducted using a 3.0 T Siemens Trio with a 12-channel phased-array head coil. For each functional run, a single-shot echo planar imaging that is sensitive to
blood oxygen level dependent (BOLD) contrast was used to acquire 33 slices (3 mm thickness) per repetition time (TR), with a TR of 2000 ms, an echo time of 30 ms, a flip angle of 90
degrees, a field of view of 192 mm, and a 64 $\times$ 64 acquisition matrix. Image preprocessing was performed using the FMRIB (Oxford Centre for Functional Magnetic Resonance Imaging of the Brain) Software Library (FSL), and motion correction was performed in MCFLIRT (Motion Correction using FMRIB's Linear Image Registration Tool). Images were high-pass filtered with a 50 s cutoff period. Spatial smoothing was performed using a kernel where full width at half maximum was 8 mm. Signals were normalized globally to account for transient fluctuations in intensity.

The whole brain is parcellated into a set of $N$ regions of interest that correspond to the 112 cortical and subcortical structures anatomically identified in FSL's Harvard-Oxford atlas. For each individual fMRI data set, we estimate regional mean BOLD time series by averaging voxel time series in each of the $N$ regions. These regional time series are then subjected to a wavelet decomposition to reconstruct wavelet coefficients in the 0.06--0.12 Hz range (scale two). We estimate the correlation or coherence $A_{ij}$ between the activity of all possible pairs of regions $i$ and $j$ to construct $N \times N$ functional connectivity matrices $\mathbf{A}$ (see Figure 1A). Individual elements of $A_{ij}$ are subjected to statistical testing, and the value of all elements that do not pass the false discovery rate correction for multiple comparisons are set to zero; otherwise, the values remain unchanged. The complete set of weighted network nodes is partitioned into communities by maximizing the modularity index $Q$ with respect to the connectivity of a random null model\cite{Porter2009,Fortunato2010}. In the simplest static case, supposing that node $i$ is assigned to community $g_{i}$ and node $j$ is assigned to community $g_{j}$, then $Q$ is defined as
\begin{equation}
	Q = \sum_{ij} [A_{ij} - P_{ij}] \delta(g_{i},g_{j})\,,
\end{equation}
where $\delta(g_{i},g_{j})=1$ if $g_{i} = g_{j}$ and it equals $0$ otherwise, and $P_{ij}$ is the expected weight of the edge connecting node $i$ and node $j$ under a specified null model. (Note: a more complex formula is used in the dynamic network case; see Supplementary Information.) The elements of the matrix $A_{ij}$ are weighted by the functional association between regions and we thoroughly sample the distribution of partitions that provide near-optimal $Q$ values \cite{Good2010}. The functional connectivity is termed `modular' if the value of $Q$ is larger than that expected from random network null models that control for both the mean and variability of connectivity.

We tested for static modular structure on these individual networks and on dynamic network structure on a multi-network object created by linking networks between time steps \cite{Mucha2010}. In both cases, we assess modular organization using the modularity $Q$ and the number of modules $n$. In the dynamic case, we also used two additional diagnostics to characterize modular structure including the mean module size $s$ and the stationarity of modules $\zeta$. We defined $s$ to be the mean number of nodes per community over all time windows over which the community exists. We used the definition of module stationarity from Ref.~\cite{Palla2007}.  We started by calculating the autocorrelation function $U(t)$ of two states of the same community $G(t)$ at $t=1$ time steps apart using the formula
\begin{equation}
	U(t) \equiv \frac{|G(t_{0})\cap G(t_{0}+t)|}{|G(t_{0})\cup G(t_{0}+t)|}\,,
\end{equation}
where $t_{0}$ is the time at which the community is born, $|G(t_{0})\cap G(t_{0}+t)|$ is the number of nodes that are members of both $G(t_{0})$ and $G(t_{0}+t)$, and $|G(t_{0})\cup
G(t_{0}+t)|$ is the total number of nodes in $G(t_{0}) \cup G(t_{0}+t)$ \cite{Palla2007}. We defined $t'$ to be the final time step before the community is extinguished. The stationarity
of a community is then
\begin{equation}
	\zeta \equiv \frac{\sum_{t=t_{0}}^{t'-1} U(t,t+1)}{t'-t_{0}-1}\,,
\end{equation}
which is the mean autocorrelation over consecutive time steps \cite{Palla2007}.

In principle, modular architecture might vary with learning by displaying changes in global diagnostics such as the number of modules or the modularity index $Q$ or by displaying more specific changes in the composition of modules. To measure changes in the composition of modules, we defined the \emph{flexibility} of a node $f_{i}$ to be the number of times that a node changed modular assignment throughout the session, normalized by the total number of changes that were possible (i.e., by the number of consecutive pairs of layers in the multilayer framework).  We then defined the flexibility of the entire network as the mean flexibility over all nodes in the network: $F = \frac{1}{N} \sum_{i=1}^{N} f_{i}$.

See Supplementary Materials for further mathematical details and methodological descriptions.
\end{materials}


\begin{acknowledgments}
This work was supported by the David and Lucile Packard Foundation, PHS Grant NS44393, the Institute for Collaborative Biotechnologies through contract no. W911NF-09-D-0001 from the U.S. Army Research Office, and the NSF (DMS-0645369).  M.A.P. acknowledges a research award (\#220020177) from the James S. McDonnell Foundation. We thank Aaron Clauset for useful discussions and John Bushnell for technical support.
\end{acknowledgments}






\end{article}



\begin{figure}
\begin{center}
\includegraphics[width=.5\textwidth]{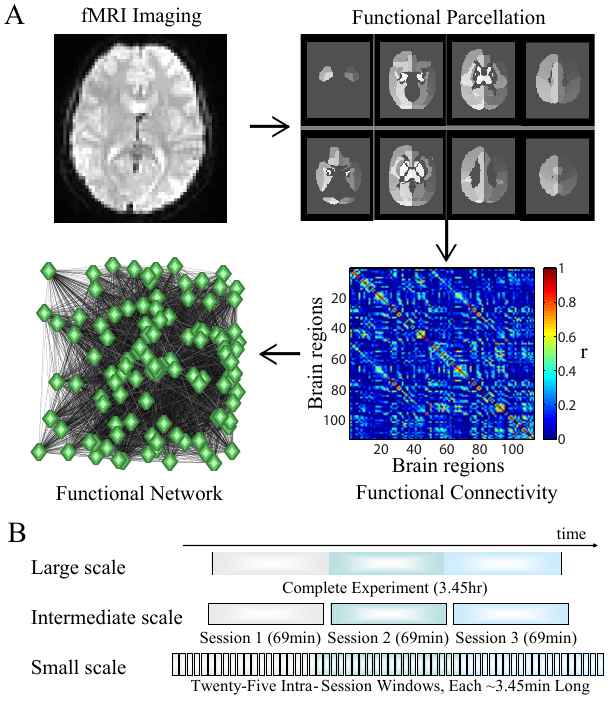}
\caption{\textbf{Structure of the Investigation} \emph{(A)} To characterize the network structure of low-frequency functional connectivity \cite{Bullmore2009} at each temporal scale, we partitioned the raw fMRI data (left, top) from each subject's brain into signals originating from $N=112$ cortical structures, which constitute the network's `nodes' (right, top).  The functional connectivity, constituting the network `edges', between two cortical structures is given by a Pearson correlation between the mean regional activity signals (right, middle).  We then statistically corrected the resulting $N\times N$ correlation matrix using a false discovery rate correction \cite{Genovese2002} to construct a subject-specific weighted functional brain network (left, middle). \emph{(B)} Schematic of the investigation that was performed over the temporal scales of days, hours, and minutes.  The complete experiment, which defines the largest scale, took place over the course of three days. At the intermediate scale, we conducted further investigations of the experimental sessions that occurred on each of those three days. Finally, to examine higher-frequency temporal structure, we cut each experimental session into 25 non-overlapping windows, each of which was a few minutes in duration. \label{Fig0}}
\end{center}
\end{figure}

\begin{figure*}
\begin{center}
\includegraphics[width=1\textwidth]{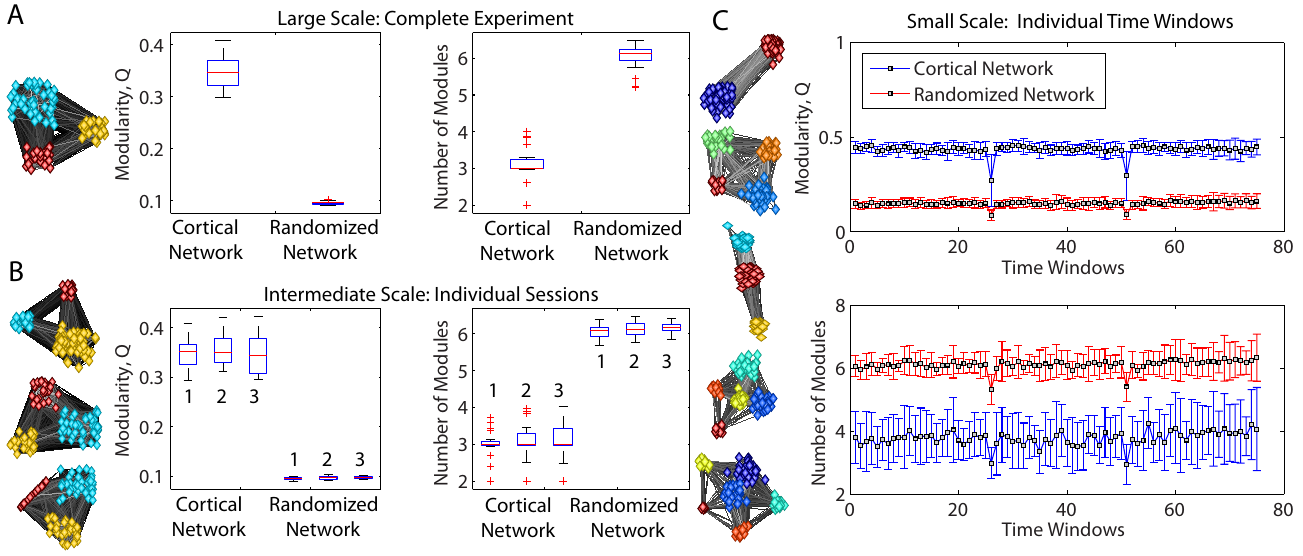}
\caption{\textbf{Multiscale Modular Architecture} \emph{(A)} Results for the modular decomposition of functional connectivity across temporal scales. In each panel, the network plots on the left show the extracted modules; different colors indicate different modules. Panels \emph{(A)} and \emph{(B)} correspond to the entire experiment and individual sessions, respectively. Boxplots show the modularity index $Q$ (left) and the number of modules (right) in the brain network compared to randomized networks. See Methods for a formal definition of $Q$. Panel \emph{(C)} shows $Q$ and the number of modules for the cortical (blue) compared to randomized networks (red) over the 75 time windows. Error bars indicate standard deviation in the mean over subjects. \label{Fig1}}
\end{center}
\end{figure*}

\begin{figure}
\begin{center}
\includegraphics[width=.5\textwidth]{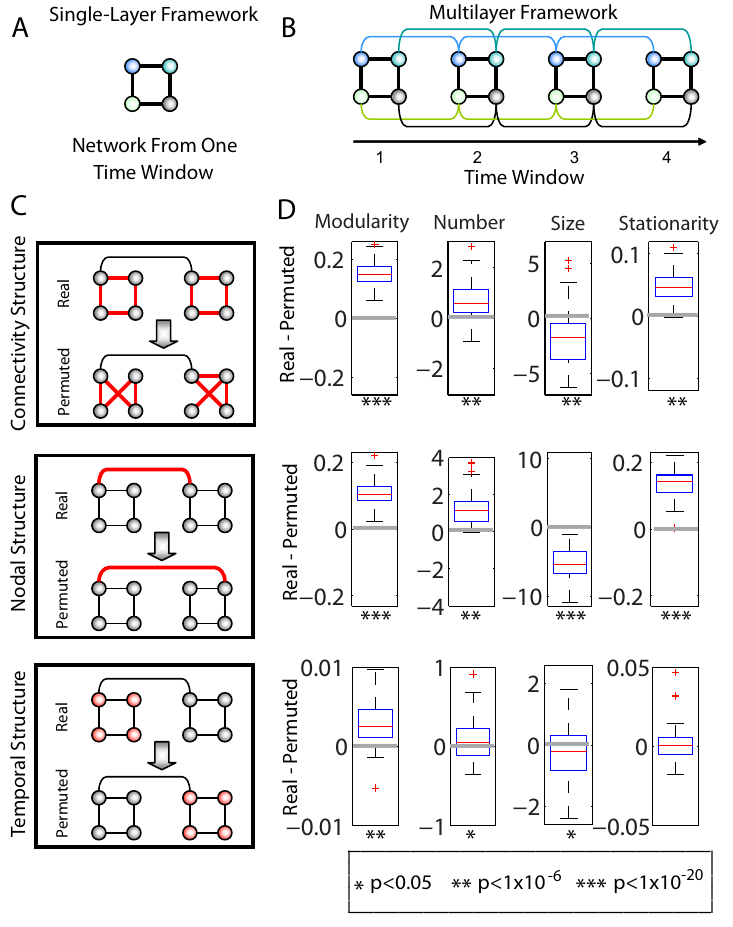}
\caption{\textbf{Temporal Dynamics of Modular Architecture} \emph{(A)} Schematic of a toy network with four nodes and four edges in a single time window. \emph{(B)} Multilayer network framework in which the networks from four time windows are linked by connecting nodes in one time window to themselves in the adjacent time windows (colored curves). \emph{(C)} Statistical framework composed of a connectional null model (top), a nodal null model (middle), and a temporal null model (bottom) in which intra-network links, inter-network links, and time windows respectively (shown in red) in the real network are randomized in the permuted network. \emph{(D)} Boxplots showing differences in modular architecture between the real and permuted networks for the connectional (top), nodal (middle), and temporal (bottom) null models. We measured the structure of the network using the modularity index $Q$, the number of modules, the module size, and stationarity, which is defined as the mean similarity in the nodal composition of modules over consecutive time steps.  Below each plot, we indicate by asterisks the significance of one-sample $t$-tests that assess whether the differences that we observed were significantly different from zero (gray lines): a single asterisk indicates $p<.05$, two asterisks indicate $p<1 \times 10^{-6}$, and three asterisks indicate $p<1 \times 10^{-20}$. \label{Fig2}}
\end{center}
\end{figure}

\begin{figure}
\begin{center}
\includegraphics[width=.5\textwidth]{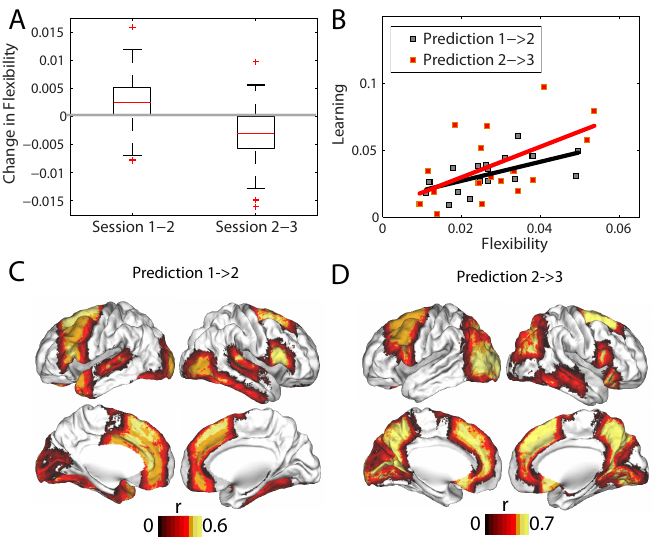}
\caption{\textbf{Flexibility and Learning} \emph{(A)} Boxplots showing that the increase in flexibility from experimental Session 1 to Session 2 was significantly greater than zero (a one-sample $t$-test gives the result $t \approx 6.00$ with $p \approx 2 \times 10^{-8}$), and that the magnitude of the decrease in flexibility from Session 2 to Session 3 was significantly greater than zero ($t \approx 7.46$, $p \approx 2 \times 10^{-11}$). \emph{(B)} Significant predictive correlations between flexibility in Session 1 and learning in Session 2 (black curve; $p \approx 0.001$) and between flexibility in Session 2 and learning in Session 3 (red curve; $p \approx 0.009$). Note that relationships between learning and network flexibility in the same experimental sessions (1 and 2) were not significant; we obtained $p > 0.13$ using permutation tests. \emph{(C)} Brain regions whose flexibility in Session 1 predicted learning in Session 2 ($p<0.05$; uncorrected for multiple comparisons). Regions that also passed false-positive correction were the left anterior fusiform cortex and the right inferior frontal gyrus, thalamus, and nucleus accumbens. \emph{(D)} Brain regions whose flexibility in Session 2 predicted learning in Session 3 ($p<0.05$; uncorrected for multiple comparisons). Regions that also passed false-positive correction for multiple comparisons were the left intracalcarine cortex, paracingulate gyrus, precuneus, and lingual gyrus and the right superior frontal gyrus and precuneus cortex. In panels \emph{(C-D)}, color indicates the Spearman correlation coefficient $r$ between flexibility and learning. \label{Fig3}}
\end{center}
\end{figure}






\end{document}


\maketitle

~\\
$^{1}$Complex Systems Group, Department of Physics, University of California, Santa Barbara, CA 93106, USA \\
$^{2}$Department of Psychology and UCSB Brain Imaging Center, University of California, Santa Barbara, CA 93106, USA \\
$^{3}$Oxford Centre for Industrial and Applied Mathematics, Mathematical Institute, University of Oxford, Oxford OX1 3LB, UK \\
$^{4}$CABDyN Complexity Centre, University of Oxford, Oxford OX1 1HP, UK \\
$^{5}$Carolina Center for Interdisciplinary Applied Mathematics, Department of Mathematics, University of North Carolina at Chapel Hill, NC 27599, USA \\
$^{6}$Institute for Advanced Materials, Nanoscience \& Technology, University of North Carolina, Chapel Hill, NC 27599, USA \\


\newpage
\tableofcontents

\newpage
\newpage



\section*{Full Description of Methods}
\addcontentsline{toc}{section}{Full Description of Methods}


\subsection*{Sample}
\addcontentsline{toc}{subsection}{Sample}

Twenty-five right-handed participants (16 female, 9 male) volunteered with informed consent in accordance with the
Institutional Review Board/Human Subjects Committee, University
of California, Santa Barbara. Handedness was determined by the Edinburgh Handedness Inventory. The mean age of the participants was 24.25 years (range 18.5--30 years). Of these, 2 participants were
removed because their task accuracy was less than 60\% correct, 1 was removed because of a
cyst in presupplementary motor area (preSMA), and 4 were removed for
shortened scan sessions.  This left 18 participants in
total. All participants had less than 4 years of experience
with any one musical instrument, had normal vision, and had no
history of neurological disease or psychiatric disorders.
Participants were paid for their participation. All
participants completed 3 training sessions in a 5-day
period, and each session was performed inside the Magnetic Resonance Imaging (MRI) scanner.


\subsection*{Experimental Setup and Procedure}
\addcontentsline{toc}{subsection}{Experimental Setup and
Procedure}

\begin{figure}
\begin{center}
\includegraphics[width=1\textwidth]{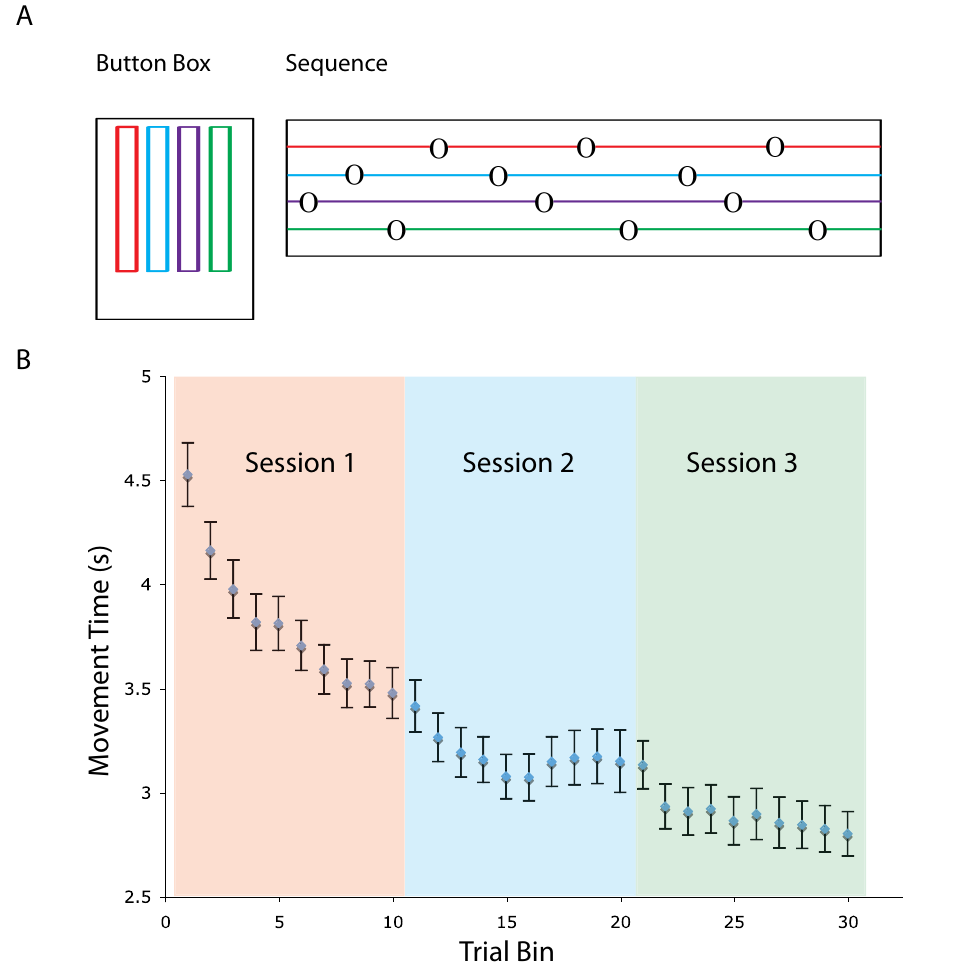}
\caption{\textbf{Experimental Setup and Learning} \emph{(A)} Schematic of the cued sequence production (CSP) task. The response or ``button'' box (left) had four response buttons that were color-coded to match the notes on the ``musical staff"  (right) presented to the subject in the visual stimulus. This visual stimulus was composed of 12 notes in sequence.  Here we show one example of a single sequence. \emph{(B)} Movement time as a function of practiced trials, whose decreasing slope indicates that learning is occuring. (We have aggregated trials into 10 trial bins per session.) \label{Fig1}}
\end{center}
\end{figure}

Participants were placed in a supine position in the MRI
scanner. Padding was placed under the knees in order to
maximize comfort and provide an angled surface to position the
stimulus response box. Padding was placed under the left
forearm to minimize muscle strain when participants
typed sequences. Finally, in order to minimize head motion,
padded wedges were inserted between the participant and head
coil of the MRI scanner. For all sessions, participants
performed a cued sequence production (CSP) task (see Figure S\ref{Fig1}), responding to
visually cued sequences by generating responses using their
non-dominant (left) hand on a custom fiber-optic response box.
For some participants, a small board was placed between the
response box and the lap in order to help balance the box
effectively. Responses were made using the 4 fingers of the
left hand (the thumb was excluded). Visual cues were presented as a
series of musical notes on a 4-line music staff. The notes
were reported in a manner that mapped the top line of the staff
to the leftmost key depressed with the pinkie finger and so on, so that notes found on the bottom line mapped onto the rightmost key with the index finger (Figure S\ref{Fig1}B). Each
12-element note sequence contained 3 notes per line, which
were randomly ordered without repetition and free of
regularities such as trills (e.g., 121) and runs (e.g., 123). The number and order of sequence trials was identical for all participants.

A trial began with the presentation of a fixation signal, which
was displayed for $2$ sec. The complete 12-element sequence
was presented immediately following the removal of the
fixation, and participants were then instructed to respond
as soon as possible. They were given a period of 8 sec to
type each sequence correctly. Participants trained on a set of
16 unique sequences, and there were three different levels
of training exposure.  Over the course of the three training
sessions, three sequences---known as \textit{skilled} sequences---were presented frequently, with 189 trials for each sequence. A second set of three sequences, termed \textit{familiar} sequences,
were presented for 30 trials each throughout training.  A
third set composed of 10 different sequences, known as
\textit{novice} sequences, were also presented; each novice sequence was presented 4--8 times during training.

Skilled and familiar sequences were practiced in blocks of 10 trials, so that 9 out of 10 trials were composed of the same sequence and 1 of the trials contained a novice sequence. If a sequence was reported correctly, then the notes were immediately removed from the screen and replaced with the
fixation signal, which remained on the screen until the trial
duration (8 sec) was reached. If there were any incorrect movements, then the
sequence was immediately replaced with the verbal cue
\textit{INCORRECT} and participants subsequently waited for the start
of the next trial. Trials were separated with an inter-trial interval (ITI) lasting
between 0 sec and 20 sec, not including any time remaining from
the previous trial. Following the completion of each block, feedback (lasting 12 sec
and serving as a rest) was presented that detailed the number of correct trials and the mean time that was taken to complete a sequence. Training epochs contained 40 trials (i.e., 4 blocks) and lasted a total of 345 scan repetition times (TRs), which took a total of 690 sec.  There were 6 scan epochs per training session (2070 scan TRs). In total, each skilled sequence was presented 189 times over the course of training (18 scan epochs; 6210 TRs).

In order to familiarize participants with the task, they were
given a short series of warm-up trials the day before the
initial training session inside the scanner. Practice was also
given in the scanner during the acquisition of the structural
scans and just prior to the start of the first training-session
epoch.  Stimulus presentation was controlled with MATLAB$^{\textregistered}$ version 7.1 (Mathworks, Natick, MA) in conjunction with
Cogent 2000 (Functional Imaging Laboratory, 2000). Key-press responses and response times
were collected using a fiber-optic custom button box transducer
that was connected to a digital response card (DAQCard-6024e; National
Instruments, Austin, TX). We assessed learning using the slope of the movement time (MT), which is the difference between the time of the first button press and the time of the last button press in a single sequence (see Figure S\ref{Fig1}B)\cite{Snoddy1926}. The negative slope of the movement curve over trials indicates that learning is occurring\cite{Snoddy1926}.


\subsection*{Acquisition and Preprocessing of fMRI Data}
\addcontentsline{toc}{subsection}{Acquisition and Preprocessing
of fMRI Data}

Functional MRI (fMRI) recordings were conducted using a 3.0 T Siemens
Trio with a 12-channel phased-array head coil. For each
functional run, a single-shot echo planar imaging that is sensitive to
blood oxygen level dependant (BOLD) contrast was used to
acquire 33 slices (3 mm thickness) per repetition time (TR),
with a TR of 2000 ms, an echo time (TE) of 30 ms, a flip angle of 90
degrees, and a field of view (FOV) of 192 mm. The spatial resolution of the data was defined by a 64 $\times$ 64 acquisition
matrix. Before the collection of the first functional epoch, a high-resolution T1-weighted sagittal sequence image of the
entire brain was acquired (TR = 15.0 ms, TE = 4.2 ms, flip angle
= 9 degrees, 3D acquisition, FOV = 256 mm; slice thickness =
0.89 mm, and spatial acquisition matrix dimensions = 256 $\times$ 256).

All image preprocessing was performed using the FMRIB (Oxford Centre for Functional Magnetic Resonance Imaging of the Brain) Software Library (FSL) \cite{Smith2004}. Motion correction was performed using the program MCFLIRT (Motion Correction using FMRIB's Linear Image Registration Tool). Images were high-pass filtered with a 50 sec cutoff period. Spatial smoothing was performed
using a kernel where the full width at half maximum was 8 mm. No temporal smoothing was performed. The signals were normalized globally to account for transient fluctuations in signal intensity.


\subsection*{Partitioning the Brain into Regions of Interest}
\addcontentsline{toc}{subsection}{Partitioning the Brain into Regions of Interest}
Brain function is characterized by a spatial specificity: different portions of the cortex emit inherently different activity patterns that depend on the experimental task at hand. In order to measure the functional connectivity between these different portions, it is common to apply an atlas of the entire brain to raw fMRI data in order to combine information from all 3 mm cubic voxels found in a given functionally or anatomically defined region (for recent reviews, see \cite{Bassett2006b,Bassett2009b,Bullmore2009}). Several atlases are currently available, and each provides slightly different parcellations of the cortex into discrete volumes of interest. Several recent studies have highlighted the difficulty of comparing results from network analyses derived from different atlases \cite{Wang2009,Zalesky2010,Bassett2010c}. In the present work, we have therefore used a single atlas that provides the largest number of uniquely identifiable regions---this is the Harvard-Oxford (HO) atlas, which is available through the FSL toolbox \cite{Smith2004,Woolrich2009}. The HO atlas provides 112 functionally and anatomically defined cortical and subcortical regions; for a list of the brain regions, see Supplementary Table 1. Therefore, for each individual fMRI data set, we estimated regional mean BOLD time series by averaging voxel time series in each of the 112 regions. Each regional mean time series was composed of 2070 time points for each of the 3 experimental sessions (for a total of 6210 time points for the complete experiment).


\subsection*{Wavelet Decomposition}
\addcontentsline{toc}{subsection}{Wavelet Decomposition}

Brain function is also characterized by a frequency specificity; different cognitive and physiological functions are associated with different frequency bands, which can be investigated using wavelets. Wavelet decompositions of fMRI time series have been applied extensively in both resting-state and task-based conditions \cite{Bullmore2003,Bullmore2004}. In both cases, they provide increased sensitivity for the detection of small signal changes in non-stationary time series with noisy backgrounds \cite{Brammer1998}. In particular, the maximum-overlap discrete wavelet transform (MODWT) has been extensively used in connectivity investigations of fMRI \cite{Achard2006,Bassett2006a,Achard2007,Achard2008,Bassett2009,Lynall2010}.  Accordingly, we used MODWT to decompose each regional time series into wavelet scales corresponding to specific frequency bands \cite{Percival2000}.  We were interested in quantifying high-frequency components of the fMRI signal, correlations between which might be indicative of cooperative temporal dynamics of brain activity during a task. Because our sampling frequency was 2 sec (1 TR = 2 sec), wavelet scale one provided information on the frequency band 0.125--0.25 Hz and wavelet scale two provided information on the frequency band 0.06--0.125 Hz. Previous work has indicated that functional associations between low-frequency components of the fMRI signal (0--0.15 Hz) can be attributed to task-related functional connectivity, whereas associations between high-frequency components (0.2--0.4 Hz) cannot \cite{Sun2004}. This frequency specificity of task-relevant functional connectivity is likely to be due at least in part to the hemodynamic response function, which might act as a noninvertible bandpass filter on underlying neural activity \cite{Sun2004}. In the present study, we therefore restricted our attention to wavelet scale two in order to assess dynamic changes in task-related functional brain architecture over short time scales while retaining sensitivity to task-perturbed endogenous activity \cite{Barnes2009}, which is most salient at about 0.1 Hz \cite{Lowe1998,Cordes2000,Cordes2001}.


\subsection*{Connectivity Over Multiple Temporal Scales}
\addcontentsline{toc}{subsection}{Connectivity over Multiple Temporal Scales}


\paragraph{Multiscale Connectivity Estimation}

We measured functional connectivity over three temporal
scales: the large scale of the complete experiment
(which lasted 3 hours and 27 minutes), the session time scale of each
fMRI recording session (3 sessions of 69 minutes
each; each session corresponded to 2070 time points), and the shorter time scales of intra-session time windows (where each time window was approximately 3.5 min long and lasted
80 time points).

In the investigation of large-scale connectivity, we concatenated regional mean time series over all 3 sessions, as has been done previously \cite{Fair2007}. We then constructed for each subject a functional association matrix based on correlations between regional mean time series. At the mesoscopic scale, we extracted regional mean time series from each experimental session separately to compute session-specific matrices. At the small scale, we constructed intra-session time windows with a length of $T=80$ time points, giving a total of 25 time windows in each session (see the Results section of this supplementary document for a detailed investigation across a range of $T$ values). We constructed separate functional association matrices for each subject in each time window (25) for each session (3) for a total of 75 matrices per subject.  We chose the length of the time window to be long enough to allow adequate estimation of correlations over the frequencies that are present in the wavelet band of interest (0.06--0.12 Hz), yet short enough to allow a fine-grained measurement of temporal evolution over the full experiment.


\paragraph{Construction of Brain Networks}

To construct a functional network, we must first define a measure of functional association between regions. Measures of functional association range from simple linear correlation to nonlinear measures such as mutual information. In the majority of network investigations in fMRI studies to date, the measure of choice has been the Pearson correlation\cite{Achard2006,Achard2007,Meunier2008,Meunier2009,Lynall2010}, perhaps due to its simplicity and ease of interpretation. Therefore, in order to estimate static functional association, we calculated the Pearson correlation between the regional mean time series of all possible pairs of regions $i$ and $j$.  This yields an $N\times N$ correlation matrix with elements $r_{i,j}$, where $N = 112$ is the number of brain regions of interest in the full brain atlas (see earlier section on ``Partitioning the Brain into Regions of Interest'' for further details).

However, as pointed out in other network studies of fMRI data\cite{Achard2006}, not all elements $r_{i,j}$ of the full correlation matrix necessarily indicate significant functional relationships. Therefore, in addition to the correlation matrix element $r_{i,j}$, we computed the $p$-value matrix element $p_{i,j}$, which give the probabilities of obtaining a correlation as large as the observed value $r_{i,j}$ by random chance when the true correlation is zero. We estimated $p$-values using approximations based on the $t$-statistic using the MATLAB$^{\textregistered}$ function {\tt corrcoef}\cite{Rahman1968}. In the spirit of Ref.~\cite{He2007} and following Ref.~\cite{Achard2006}, we then tested the $p$-values $p_{i,j}$ for significance using a False Discovery Rate (FDR) of $p < 0.05$ to correct for multiple comparisons\cite{Benjamini2001,Genovese2002}.  We retained matrix elements $r_{i,j}$ whose $p$-values $p_{i,j}$ passed the statistical FDR threshold. Elements of $r_{i,j}$ whose $p$-values $p_{i,j}$ did not pass the FDR threshold were set to zero in order to create new correlation matrix elements $r'_{i,j}$.

We applied the statistical threshold to all $r_{i,j}$ independent of the sign of the correlation. Therefore, the resulting $r'_{i,j}$ could contain both positive and negative elements if there existed both positive and negative elements of $r_{i,j}$ whose $p$-values $p_{i,j}$ passed the FDR threshold. Because this was a statistical threshold, the network density of $r'_{i,j}$ (defined as the fraction of non-zero matrix elements) was determined statistically rather than being set \textit{a priori}. Network density varied over temporal resolutions; the mean density and standard deviation for networks derived from correlation matrices at the largest time scale (3 hr and 27 minutes) was 0.906 (0.019\%), at the intermediate time scale (69 min) was 0.846 (0.029), and at the short time scale (3.5 min) was 0.423 (0.110).

We performed the procedure described above for each subject separately to create subject-specific corrected correlation matrices. These statistically corrected matrices gave adjacency matrices $\mathbf{A}$ (see the discussion below) whose elements were $A_{ij} = r'_{i,j}$.


\paragraph{Network Modularity}

To characterize the large-scale functional organization of the subject-specific weighted matrices $\mathbf{A}$, we used tools from network science \cite{Newman2010}.  In a network framework, brain regions constitute the \textit{nodes} of the network, and inter-regional functional connections that remain in the connectivity matrix constitute the \textit{edges} of the network.  One powerful concept in the study of networks is that of \emph{community structure}, which can be studied using algorithmic methods \cite{Porter2009,Fortunato2010}.  Community detection is an attempt to decompose a system into subsystems (called `modules' or `communities').  Intuitively, a module consists of a group of nodes (in our case, brain regions) that are more connected to one another than they are to nodes in other modules.  A popular way to investigate community structure is to optimize the partitioning of nodes into modules such that the \emph{quality function} $Q$ is maximized (see \cite{Porter2009,Fortunato2010} for recent reviews and \cite{Good2010} for a discussion of caveats), for which we give a formula below.

From a mathematical perspective, the quality function $Q$ is simple to define. One begins with a graph composed of $N$ nodes and some set of connections between those nodes. The adjacency matrix $\mathbf{A}$ is
then an $N\times N$ matrix whose elements $A_{ij}$ detail a
direct connection or `edge' between nodes $i$ and $j$, with a
weight indicating the strength of that connection. The quality of a hard partition of $\mathbf{A}$ into communities (whereby each node is assigned to exactly one community) is then quantified using the quality function $Q$. Suppose that node $i$ is assigned to community
$g_{i}$ and node $j$ is assigned to community $g_{j}$.  The most popular form of the quality function takes the form \cite{Porter2009,Fortunato2010}
\begin{equation}
	Q = \sum_{ij} [A_{ij} - P_{ij}] \delta(g_{i},g_{j})\,,
\end{equation}
where $\delta(g_{i},g_{j})=1$ if $g_{i} = g_{j}$ and it equals $0$ otherwise, and
$P_{ij}$ is the expected weight of the edge connecting node $i$
and node $j$ under a specified null model. (The specific choice of $Q$ in Equation 1 is called the \emph{network modularity} or \emph{modularity index} \cite{Newman2002b}.)
A most common null model (by far) used for static network community detection is given by \cite{Porter2009,Fortunato2010,Girvan2002}
\begin{equation}
P_{ij} = \frac{k_{i} k_{j}}{2m}\,,
\end{equation}
where $k_{i}$ is the strength of node $i$, $k_{j}$ is the strength of node $j$, and $m=\frac{1}{2}\sum_{ij} A_{ij}$. The
maximization of the modularity index $Q$ gives a partition
of the network into modules such that the total edge weight
inside of modules is as large as possible (relative to the null model, subject to the limitations of the employed computational heuristics, as optimizing $Q$ is NP-hard \cite{Porter2009,Fortunato2010,Brandes2008}).

Network modularity has been used recently for investigations of resting-state functional brain
networks derived from fMRI \cite{Meunier2008,Meunier2009} and of anatomical brain networks derived from morphometric
analyzes \cite{Chen2008}.  In these previous studies, brain networks were constructed as undirected binary graphs, so that each edge had a weight of either $1$ or $0$.  The characteristics of binary graphs derived from neuroimaging data are sensitive to a wide variety of cognitive, neuropsychological, and neurophysiological factors \cite{Bassett2009b,Bullmore2009}.  However, increased sensitivity is arguably more likely in the context of the weighted graphs that we consider, as they preserve the information regarding the strength of functional associations (though, as discussed previously, matrix elements $r_{i,j}$ that are statistically insignificant are still set to $0$)  \cite{Rubinov2009}. An additional contrast between previous studies and the present one is that (to our knowledge) investigation of network modularity has not yet been applied to task-based fMRI experiments, in which modules might have a direct relationship with goal-directed function.

We partitioned the networks represented by the weighted connectivity matrices into $n$ communities by using a Louvain greedy community detection method \cite{Blondel2008} to optimize the modularity index $Q$. Because the edge weights in the correlation networks that we constructed contain both positive and negative correlation coefficients, we used the signed null model proposed in Ref.~\cite{Traag2009} to account for communities of nodes associated with one another through both negative and positive edge weights.  (Recall that we are presently discussing aggregated correlation networks $A$, so we are detecting communities in single-layer networks, as has been done in previous work.  In order to investigate time-evolving communities, we will later employ a new mathematical development that makes it possible to perform community detection in multilayer networks\cite{Mucha2010}.) We first defined $w_{ij}^{+}$ to be an $N\times N$ matrix containing the positive elements of $A_{ij}$ and $w_{ij}^{-}$ to be an $N\times N$ matrix containing only the negative elements of $A_{ij}$. The quality function to be maximized is then given by
\begin{equation}
	 Q_{\pm} = \frac{1}{2w^{+} + 2w^{-}} \sum_{i} \sum_{j} \left[ A_{ij} - \left(\gamma^{+} \frac{w_{i}^{+}w_{j}^{+}}{2w^{+}} - \gamma^{-} \frac{w_{i}^{-}w_{j}^{-}}{2w^{-}}\right) \right] \delta(g_{i}g_{j})\,,
\end{equation}
where $g_{i}$ is the community to which node $i$ is assigned, $g_{j}$ is the community to which node $j$ is assigned, $\gamma^{+}$ and $\gamma^{-}$ are resolution parameters, and $w_{i}^{+} = \sum_{j} w_{ij}^{+}$, $w_{i}^{-} = \sum_{j} w_{ij}^{-}$ \cite{Traag2009}. For simplicity, we set the resolution parameter values to unity.

In our investigation, we have focused on the mean properties of ensembles of partitions rather than on detailed properties of individual partitions. This approach is consistent with recent work illustrating the fact that the optimization of quality functions like $Q$ and $Q_{\pm}$ is hampered by the complicated shape of the optimization landscape. In particular, one expects to find a large number of partitions with near-optimum values of the quality function \cite{Good2010}, collectively forming a high-modularity plateau. Theoretical work estimates that the number of ``good'' (in the sense of high values of $Q$ and similar quality functions) partitions scales as $2^{\bar{n}-1}$, where $\bar{n}$ is the mean number of modules in a given partition \cite{Good2010}.  In both toy networks and networks constructed from empirical data, many of the partitions found by maximizing a quality function disagree with one another on the components of even the largest module, impeding interpretations of particular partitions of a network \cite{Good2010}. Therefore, in the present work, we have focused on quantifying mean qualities of the partitions after extensive sampling of the high-modularity plateau. Importantly, the issue of extreme near-degeneracy of quality functions like $Q$ is expected to be much less severe in the networks that we consider than is usually the case, because we are examining small, weighted networks rather than large, unweighted networks \cite{Good2010}. We further investigate the degenerate solutions in terms of their mean, standard deviation, and maximum. We find that $Q_{\pm}$ values are tightly distributed, with maximum values usually less than three standard deviations from the mean (see Supplementary Results).

\paragraph{Statistical Testing}

To determine whether the value of $Q_{pm}$ or the number of modules
was greater or less than expected in a random system, we
constructed randomized networks with the same degree
distribution as the true brain networks. As has been done previously\cite{Bassett2010,Meunier2009}, we began with a real
brain network and then iteratively rewired it using the
rewiring algorithm of Maslov and Sneppen \cite{Maslov2002}. The procedure we used for accomplishing this rewiring was to choose at random two edges---one that connects node A to node B and another that connects nodes C and D---and then to rewire them to connect A to C and B to D.  This allows us to preserve the degree, or number of edges, emanating from each node although it does not retain a node's strength.  To ensure a thorough randomization of the underlying connectivity structure, we performed this procedure multiple times, such that the expected number of times that each edge was `rewired' was 20. This null model will be hereafter referred to as the \textit{static random network null model}. (This is distinct from the null models that we have developed for statistical testing of community structure in multilayer networks, as discussed in the main manuscript and in later sections of this Supplement.) The motivation for this process is to compare the brain with a null model that resembles the configuration model \cite{Molloy1995}, which is a random graph with prescribed degree distribution.

We constructed 100 instantiations of the static random network null model for each real network that we studied.  We constructed representative values for diagnostics from the random networks by taking the mean network modularity and mean number of modules over those 100 random networks.  We then computed the difference between the representative random values and the real values for each diagnostic, and we performed a one-sample $t$-test over subjects to determine whether that difference was significantly greater than or less than zero. For each case, we then reported $p$-values for these tests.

Sampling of the static random network null model distribution is important in light of the known degeneracies of modularity (which we discuss further in the Supplementary Results section below) \cite{Good2010}. One factor that accounts for a significant amount of variation in $Q_{\pm}$ is the size (i.e., number of nodes) of the network, so comparisons between networks of different sizes must be performed with caution. Therefore, we note that all networks derived from the aforementioned null model retain both the same number of nodes and the same number of edges as the real networks under study. This constrains important factors in the estimation of $Q_{\pm}$.


\paragraph{Visualization of Networks}

We visualized networks using the software package MATLAB$^{\textregistered}$ (2007a, The MathWorks Inc., Natick, MA). Following Ref.~\cite{Traud2009}, we used the
Fruchtermann-Reingold algorithm \cite{Fruchtermann1991} to determine node
placement for a given network with respect to the extracted
communities and then used the Kamada-Kawai algorithm \cite{Kamada1988} to place
the nodes within each community.


\subsection*{Multilayer Network Modularity: Temporal Dynamics of Intra-Session Connectivity}
\addcontentsline{toc}{subsection}{Multilayer Network Modularity: Temporal Dynamics of
Intra-Session Connectivity}

In order to investigate the temporal evolution of modular architecture in human functional connectivity, we used a mutilayer network framework in which each layer consists of a network derived from a single time window. Networks in consecutive layers therefore correspond to consecutive time windows. We linked networks in consecutive time windows by connecting each node in one window to itself in the previous and in the next windows (as shown in Figure 3A-B in the main text) \cite{Mucha2010}. We constructed a multilayer network for each individual and in each of the three experimental sessions. We then performed community detection by optimizing a multilayer modularity (see the discussion below) \cite{Mucha2010} using the Louvain greedy algorithm (suitably adapted for this more general structure) on each multilayer network in order to assess the modular architecture in the temporal domain.

Our examination of static network architecture, we used the wavelet correlation to assess functional connectivity. Unfortunately, more sensitive measures of temporal association such as the spectral coherence are not appropriate over the long time scales assessed in the static investigation due to the nonstationarity of the fMRI time series\cite{Bullmore2003,Bullmore2004,Brammer1998}, and it is exactly for this reason that we have used the wavelet correlation for the investigation of aggregated (static) networks. However, over short temporal scales such as those being used to construct the multilayer networks, fMRI signals in the context of the motor learning task that we study can be assumed to be stationary\cite{Friston1999}, so spectral measures such as the coherence are potential candidates for the measurement of functional association.

In the examination of the dynamic network architecture of brain function using multilayer community detection, our goal was to measure temporal adaptivity of modular function over short temporal scales. In order to estimate that temporal adaptivity with enhanced precision, we used the magnitude squared spectral coherence (as estimated using the minimum-variance distortionless response method \cite{Benesty2006}) as a measure of nonlinear functional association between any two time series. In using the coherence, which has been demonstrated to be useful in the context of fMRI neuroimaging data \cite{Sun2004}, we were able to measure frequency-specific linear relationships between time series.

As in the static network analysis described earlier, we tested the elements of each $N \times N$ coherence matrix (which constitutes a single layer) for significance using an FDR correction for multiple comparisons. We used the original weighted (coherence values) of network links corresponding to the elements that passed this statistical test, while those corresponding to elements that did not pass the test were set to zero. In applying a community detection technique to the resulting coherence matrices, it is important to note that the coherence is bounded between $0$ and $1$. We can therefore use a multilayer quality function with an unsigned null model rather than the signed null model used in the static case described earlier.  The multilayer modularity $Q_{ml}$ is given by \cite{Mucha2010}
\begin{equation}
    Q_{ml} = \frac{1}{2\mu}\sum_{ijlr}\left\{\left(A_{ijl}-\gamma_l\frac{k_{il}k_{jl}}{2m_l}\right)\delta_{lr} + \delta_{ij}C_{jlr}\right\} \delta(g_{il},g_{jr})\,,
\end{equation}
where the adjacency matrix of layer $l$ (i.e., time window number $l$) has components $A_{ijl}$, $\gamma_l$ is the resolution parameter of layer $l$, $g_{il}$ gives the community assignment of node $i$ in layer $l$, $g_{jr}$ gives the community assignment of node $j$ in layer $r$, $C_{jlr}$ is the connection strength between node $j$ in layer $r$ and node $j$ in layer $l$ (see the discussion below), $k_{il}$ is the strength of node $i$ in layer $l$, $2\mu=\sum_{jr} \kappa_{jr}$, $\kappa_{jl}=k_{jl}+c_{jl}$, and $c_{jl} = \sum_r C_{jlr}$. For simplicity, as in the static network case, we set the resolution parameter $\gamma_{l}$ to unity and we have set all non-zero $C_{jlr}$ to a constant $C$, which we will term the `inter-layer coupling'. In the main manuscript, we report results for $C=1$. In the Supplementary Results section of this document, we investigate the dependence of our results on alternative choices for the value of $C$.


\paragraph{Diagnostics}

We used several diagnostics to characterize dynamic modular structure.  These include the multilayer network
modularity $Q_{ml}$, the number of modules $n$, the module size $s$, and the stationarity of modules $\zeta$. We defined the
size of a module $s$ to be the mean number of nodes per module over all time windows over which the community exists. We used the definition of module stationarity from Ref.~\cite{Palla2007}.  We started by calculating the autocorrelation function $U(t)$ of two states of the same community $G(t)$ at $t=1$ time steps apart using the formula
\begin{equation}
	U(t) \equiv \frac{|G(t_{0})\cap G(t_{0}+t)|}{|G(t_{0})\cup G(t_{0}+t)|}\,,
\end{equation}
where $t_{0}$ is the time at which the community is born, $|G(t_{0})\cap G(t_{0}+t)|$ is the number of nodes that are members of both $G(t_{0})$ and $G(t_{0}+t)$, and $|G(t_{0})\cup
G(t_{0}+t)|$ is the total number of nodes in the union of
$G(t_{0})$ and $G(t_{0}+t)$ \cite{Palla2007}. We defined $t'$ to be the final time step before the community is extinguished. The stationarity of a community is then given by
\begin{equation}
	\zeta \equiv \frac{\sum_{t=t_{0}}^{t'-1} U(t,t+1)}{t'-t_{0}-1}\,,
\end{equation}
which is the mean autocorrelation over consecutive time steps \cite{Palla2007}.


\paragraph{Statistical Framework}

The study of the ``modular architecture'' of a system is of little value if the system is not modular. It is therefore imperative to statistically quantify the presence or absence of modular
architecture to justify the use of community detection in a given application.
Appropriate random null models have been
developed and applied to the static network framework \cite{Bassett2010,Meunier2009}, but no such null models yet exist for the multilayer framework. We
therefore developed several null models in order to statistically test the
temporal evolution of modular structure.  We constructed three independent null models to test for (1) network structure dependent on the topological architecture of
connectivity, (2) network structure dependent on nodal identity,
and (3) network structure dependent on the temporal organization
of layers in the multilayer framework.

In the \emph{connectional null model} (1), we scrambled links between nodes
in any given time window (the entire experiment, 3.45 hr;
the individual scanning session, 69 min; or intra-session time
windows, 3.45 min) while maintaining the total
number of connections emanating from each node in the
system. To be more precise, for each layer of the multilayer network, we sampled the \textit{static random network null model} (see the discussion above in the context of static connectivity architecture) for that particular layer. That is, we reshuffled the connections within each layer separately while maintaining the original degree distribution. We then linked these connectivity-randomized layers together by coupling a node in one layer to itself in contiguous layers to create the connectional null model multilayer network, just as we connected the real layers to create the real multilayer network. In the present time-dependent context, we performed this procedure on each time window in the multilayer network, after which we applied the multilayer community detection algorithm to determine the network modularity of the randomized system.

In constructing a \emph{nodal null model} (2), we focused on the
links that connected a single node in one layer of the
multilayer framework to itself in the next and previous layers. In the null
model, the links between layers connect a node in one layer to randomly-chosen nodes in contiguous layers instead of connecting the node to itself in those layers. Specifically, in each time window $\tau_i$ (except for the final one), we randomly connected the nodes in the corresponding layer to other nodes in the next time window
($\tau_{i+1}$) such that no node in $\tau_{i}$ was connected to
more than one node in $\tau_{i+1}$.  We then connected nodes in $\tau_{i+1}$ to randomly-chosen nodes in $\tau_{i+2}$, and so on until links between all time windows had been fully randomized.

We also considered randomization of the order in which time windows were placed in the multilayer network to construct a \emph{temporal
null model} (3). In the real multilayer construction, we (of course) always placed the network
from time window $\tau_{i}$ just before the
network from time window $\tau_{i+1}$. In the temporal null
model, we randomly permuted the temporal location of the individual layer in the
multilayer framework such that the probability of any time window $\tau_{i}$ following any other
time window $\tau,~(j\neq i)$ was uniform.


\paragraph{Statistical Testing}

In both the real network and the networks derived from null models (1)--(3), it is important to adequately sample the distributions of partitions meant to optimize the modularity index $Q_{ml}$.  This step in our investigation was particularly important in light of the extreme degeneracy of the network modularity function $Q_{ml}$ \cite{Good2010} (see the Supplementary Results section on the Degeneracies of $Q$ for a quantitative characterization of such degeneracies).

Because the multilayer community detection algorithm can find
different maxima each time it is run, we computed the community structure
of each individual real multilayer network a total of
100 times.  We then averaged the values of all diagnostics (modularity index $Q_{ml}$,
number of modules, module size, and stationarity) over those 100 partitions to create a
representative real value.
To perform our sampling for the null models, we considered 100 multilayer network instantiations for each of the three different null models.  We also performed community detection on these null models using our multilayer network adaptation of the Louvain modularity-optimization algorithm \cite{Blondel2008} to create a distribution of
values for each diagnostic.
We then used the mean value of each diagnostic in our subsequent
investigation as the representative value of the null model.

We used one-sample $t$-tests to test statistically whether
the differences between representative values from the real
networks and null model networks over the subject population
was significantly different from zero.  The results, which we reported in the main manuscript, indicated that in contrast to what we observed using each of the three null models, the human brain displayed a heightened modular structure. That is, it is composed of more modules, which have smaller sizes. Considering the three null models in order, this suggests that cortical connectivity has a precise topological organization, that cortical regions consistently maintain individual connectivity signatures necessary for cohesive community organization, and that functional communities evolve cohesively in time (see Figure 2 in the main manuscript). Importantly, the stationarity of modular organization $\zeta$ was also higher in the human brain than in the connectional or nodal null models, indicating a cohesive temporal evolution of functional communities.


\subsection*{Temporal Dynamics of Brain Architecture and Learning}
\addcontentsline{toc}{subsection}{Temporal Dynamics of Brain Architecture and Learning}

In the present study, we have attempted to determine whether changes in the dynamic modular architecture of functional connectivity is shaped by learning. We assessed the learning in each session using the slope of the movement times (MT) of that session. Movement time is defined as the difference between the time of the first button press and the time of the last button press in a single sequence (see Figure S\ref{Fig1}B). During successful learning, movement time is known to fall logarithmically with time\cite{Snoddy1926}. However, two subjects from session 1 and one subject from session 2 showed an \emph{increasing} movement time as the session progressed. We therefore excluded these three data points in subsequent comparisons due to the decreased likelihood that successful learning was taking place. This process of screening participants based on movement time slope is consistent with previous work suggesting that fMRI activation patterns during successful performance might be inherently different when performance is unsuccessful \cite{Callicott1999}.

In principle, modular architecture might vary with learning by displaying changes in global diagnostics such as the number of modules or the modularity index $Q$ or by displaying more specific changes in the composition of modules. To measure changes in the composition of modules, we defined the \emph{flexibility} of a node $f_{i}$ to be the number of times that node changed modular assignment throughout the session, normalized by the total number of changes that were possible (i.e., by the number of consecutive pairs of layers in the multilayer framework).  We then defined the flexibility of the entire network as the mean flexibility over all nodes in the network: $F = \frac{1}{N} \sum_{i=1}^{N} f_{i}$.


\subsection*{Statistics and Software}
\addcontentsline{toc}{subsection}{Statistics and Software}

We implemented all computational and simple statistical operations using the software packages MATLAB$^{\textregistered}$ (2007a, The MathWorks Inc., Natick, MA) and Statistica$^{\textregistered}$ (version 9, StatSoft Inc.).  We performed the network calculations using a combination of in-house software (including multilayer community detection code \cite{Mucha2010}) and the Brain Connectivity Toolbox \cite{Rubinov2009}.


\newpage
\newpage

\section*{Supplementary Results}
\addcontentsline{toc}{section}{Supplementary Results}


\subsection*{Degeneracies of $Q$}
\addcontentsline{toc}{subsection}{Degeneracies of $Q$}

As discussed earlier in the Methods section, we focused in this investigation on the mean properties of ensembles of partitions rather than on detailed properties of individual partitions.  Our approach was motivated by recent work indicating that the optimization of modularity and similar quality functions is hampered by the complicated shape of the optimization landscape, which includes a large number of partitions with near-optimum values that collectively form a high modularity plateau \cite{Good2010}. To quantify and address this degeneracy of $Q_{\pm}$ and $Q_{ml}$, we now provide supplementary results on the mean, standard deviation, and maximum values of $Q_{\pm}$ and $Q_{ml}$ over the $100$ samples of the plateau computed for all real networks in both the static and dynamic frameworks.

\begin{figure}
\begin{center}
\includegraphics[width=1\textwidth]{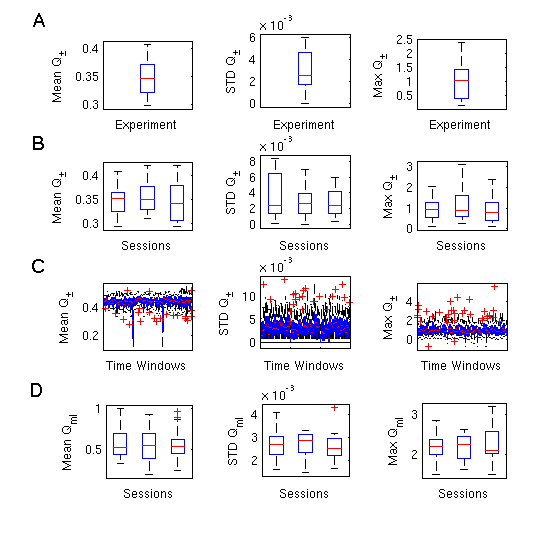}
\caption{\textbf{Properties of the static and dynamic modularity indices $Q_{\pm}$ and $Q_{ml}$}. The mean (column 1), standard deviation (column 2), and maximum (column 3) of the static modularity index $Q_{\pm}$ is shown for \emph{(A)} the large scale (entire experiment), \emph{(B)} the mesoscopic scale (three experimental sessions), and \emph{(C)} the small scale (individual time windows) over the $100$ samplings. Row \emph{(D)} shows the mean (column 1), standard deviation (column 2), and maximum (column 3) of the dynamic modularity index $Q_{ml}$ over the $100$ samplings. In the figure, the standard deviation is abbreviated as STD. Boxplots indicate 95\% confidence intervals over subjects. \label{Fig2}}
\end{center}
\end{figure}

The mean number of modules in a given partition in the static framework was $\bar{n} \approx 3.08$ for the entire experiment, $\bar{n} \approx 3.07$ for individual experimental sessions, and $\bar{n} \approx 3.55$ for the small intra-session time windows. The mean number of modules in a given partition in the multilayer framework was $\bar{n} \approx 6.00$. We have therefore chosen to sample the quality functions $Q_{\pm}$ and $Q_{ml}$ a total of $100$ times (which is more than $2^{\bar{n}-1}$ in each case, and therefore adequately samples the degenerate near-optimum values of $Q_{\pm}$ and $Q_{ml}$ \cite{Good2010}). In order to characterize the distribution of solutions found in these $100$ samplings, we have computed the mean, standard deviation, and maximum of $Q_{\pm}$ (static cases) and $Q_{ml}$ (dynamic cases); see Figure S\ref{Fig2}.  We found that the values of $Q_{\pm}$ and $Q_{ml}$ are tightly distributed, and that the maximum values of $Q_{\pm}$ or $Q_{ml}$ are between $0$ and $3$ standard deviations higher than the mean.  Although we remain cautious because we have not explored all possible computational heuristics, we are nevertheless encouraged by these results that the mean values of $Q_{\pm}$ and $Q_{ml}$ that we have reported are representative of the true maximization of the two quality functions.


\paragraph{Reproducibility}

We calculated the intra-class correlation coefficient (ICC), to determine whether values of $Q_{\pm}$ and $Q_{ml}$ derived from a single individual over the $100$ samples were more similar than values of $Q_{\pm}$ or $Q_{ml}$ derived from different individuals. The ICC is a measure of the total variance for which between-subject variation accounts \cite{Lachin2004,McGraw1996}, and it is defined as
\begin{equation}
 	\mathrm{ICC} = \frac{\sigma^{2}_{bs}}{\sigma^{2}_{bs}+\sigma^{2}_{ws}}\,,
\end{equation}
where $\sigma_{bs}$ is the between-subject variance and $\sigma_{ws}$ is the pooled within-subject variance (`pooled' indicates that variance was estimated for each subject and then averaged over subjects). The ICC is normalized to have a maximum value of $1$; values above $0.5$ indicate that there is more variability between $Q_{\pm}$ and $Q_{ml}$ values from different subjects than between $Q_{\pm}$ and $Q_{ml}$ values from the same subject. In the static framework, the ICC was $0.9884$ at the large scale (the entire experiment), an average of $0.9863$ at the intermediate scale (three experimental sessions), and an average of $0.9847$ at the small scale (individual time windows). In the multilayer framework, we calculated that ICC $\approx 0.9983$. These results collectively indicate that the $Q_{\pm}$ and $Q_{ml}$ values that we reported in this work were significantly reproducible over the $100$ samples of the respective quality function landscape. That is, the $Q_{\pm}$ or $Q_{ml}$ values drawn from the $100$ samples of a single subject's network modularity landscape were more similar than $Q_{\pm}$ or $Q_{ml}$ values drawn from different subjects.


\subsection*{Effect of the Inter-Layer Coupling Parameter}
\addcontentsline{toc}{subsection}{Effect of Inter-Layer Coupling Parameter}

The multilayer network framework requires one to define a coupling parameter $C$ that indicates the strength of the connections from a node in one time window to itself in the two neighboring time windows \cite{Mucha2010}. In order to be sensitive to both temporal dynamics and intra-layer network architecture, the coupling parameter should be on the same scale of values as the edge weights. For example, if edge weights are coherence values lying between $0$ and $1$, then the coupling parameter also ought to lie between $0$ and $1$. In the results that we presented in the main manuscript, we set the coupling parameter to be $C=1$, which is the highest value consistent with the intra-layer edge weights given by the normalized coherence. However, if we were to alter the coupling value, one might expect the number of communities to be altered in kind. As the strength of the coupling is increased, one might expect fewer communities to be uncovered due to the increased temporal dependence between layers \cite{Mucha2010}. Similarly, as the inter-layer coupling is weakened, one might expect more communities to be detected.

\begin{figure}
\begin{center}
\includegraphics[width=.45\textwidth]{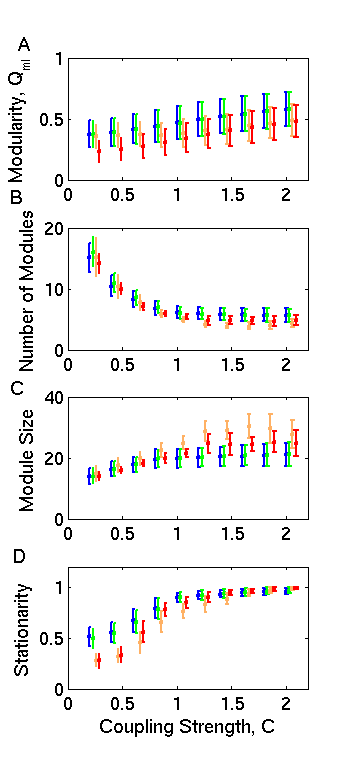}
\caption{\textbf{Effects of the coupling parameter $C$} on the four diagnostics in this study: modularity index $Q_{ml}$, number of modules $n$, module size (i.e., number of nodes) $s$, and module stationarity $\zeta$.  We first averaged values over 100 `optimal' partitions (see the discussion in the text), so this figure gives mean values of all diagnostics. The error bars indicate standard deviations over subjects and sessions. Colors indicate network type: cortical network (blue), temporal null model network (green), nodal null model network (orange), and connectional null model network (red). Error bars for different network types at a given value of $C$ (0.2, 0.4, 0.6, 0.8, 1, 1.2, 1.4, 1.6, 1.8, 2) are offset from each other for better visualization.\label{Fig3}}
\end{center}
\end{figure}

To probe the effect of the inter-layer coupling strength, we thus varied $C$ from well below to well above the maximum intra-layer edge weight ($0.2\leq C \leq2$).  In Figure S\ref{Fig3} (cortical network results are shown in blue), we illustrate the effects of sweeping over this coupling parameter on our four diagnostics.  The modularity index $Q_{ml}$ increases with increasing inter-layer coupling, whereas the other three diagnostics---number of modules, module size, and stationarity---increase initially and then plateau approximately at about $C=1$ and above. The change in behavior near $C=1$ can be rationalized as follows: For $C<1$, intra-layer edge weights dominate the modularity optimization, whereas inter-layer edge weights dominate for $C>1$. The proposed choice of $C=1$ therefore balances the impact of known coherence in brain activity (as given by the intra-layer edge weights) on measured architectural adaptations and is therefore a natural choice with which to investigate biologically meaningful organization.

We also computed $100$ temporal, nodal, and connectional null model networks for each of the additional coupling parameter values (see Figure S\ref{Fig3}; null model network results shown in green, orange, and red). The results indicate that the relationship between diagnostics in the cortical networks and null model networks is dependent on the diagnostic. For example, modularity values of null model networks are consistently lower than modularity values of cortical networks. However, stationarity in the null model networks is lower than that in cortical networks for small values of $C$ but higher than that of cortical networks for high values of $C$. This nontrivial behavior suggests an added sensitivity of the proposed null model networks to the multilayer network construction, which might be useful in other experimental contexts and therefore warrants further investigation.


\subsection*{Effect of the Time Window Length}
\addcontentsline{toc}{subsection}{Effect of the Time Window Length}

In the construction of networks at the smallest time scale, it is necessary to choose a length of the time window $T$. In choosing this time window length, two considerations are important: (1) the time window must be short enough to adequately measure temporal evolution of network structure, and (2) the time window must be long enough to adequately estimate the functional association between two time series using (for example) the correlation or coherence \cite{Fenn2009}. In the main text, we reported results for time windows of 80 data points in length.  This gives 25 time windows in each experimental session, for a total of 75 time windows over the 3 sessions. In addition to this extensive coverage of the underlying temporal dynamics, the choice of a time window of 80 data points in length also ensures that 20 data points can be used for the estimation of the functional association between time series in the frequency band of interest---i.e., at wavelet scale two (0.06--0.12 Hz). If one were to increase the time window length, one would expect a decreased ability to measure temporal variations due to the presence of fewer time windows per session. If one were to decrease the time window length, one would expect increased variance in the estimation of the functional association between time series due to the use of fewer data points in the estimation of either the coherence or the correlation \cite{Achard2008}.

\begin{figure}
\begin{center}
\includegraphics[width=.45\textwidth]{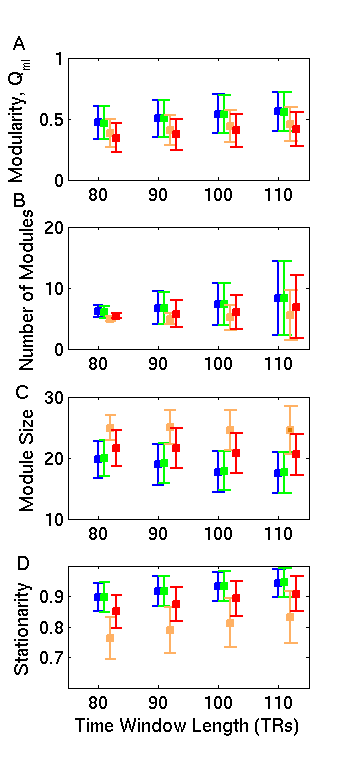}
\caption{\textbf{Effect of the time window length $T$} on the four diagnostics in this study: modularity index $Q_{ml}$, number of modules $n$, module size (i.e., number of nodes) $s$, and module stationarity $\zeta$.  We first averaged values over 100 `optimal' partitions (see the discussion in the text), so this figure gives mean values of all diagnostics. The error bars indicate standard deviations over subjects and sessions.  In the figure, we give time windows in terms of the number of data points in the time series (i.e., the number of TRs). Colors indicate network type: cortical network (blue), temporal null model network (green), nodal null model network (orange), and connectional null model network (red). Error bars for different network types at a given value of $T$ (80, 90, 100, 110) are offset from each other for better visualization. \label{Fig4}}
\end{center}
\end{figure}

To probe the effect of the time window length, we varied $T$ from $T=80$ to $T=110$ (see Figure S\ref{Fig4}; cortical network results are shown in blue). We find that the stationarity of the modules increases with increasing time window length. As $T$ is increased, the functional association between any two nodes is averaged over a longer time series, so small adaptations over shorter time scales can no longer be measured. This smoothing is likely the cause of the increased stationarity that we find at high values of $T$.  It suggests that functional association measured over long time windows is less dependent on the time window being used than functional association measured over short time windows. This finding supports our choice of short time windows in order to measure dynamic adaptations in network architecture.

We also computed $100$ temporal, nodal, and connectional null model networks for each of the additional time window lengths (see Figure S\ref{Fig4}; null model network results are shown in green, orange, and red). The results indicate that the relationships between diagnostics in the cortical networks and null model networks are largely conserved across time window lengths.


\subsection*{Learning and Flexibility}
\addcontentsline{toc}{subsection}{Learning and Flexibility}

In the main text, we reported a significant correlation between the flexibility of dynamic modular architecture in a given experimental session, as measured by the (normalized) number of times a node changes module allegiance, and learning in the subsequent experimental sessions, as measured by the slope of the movement time (see Methods). We found that the mean value of flexibility was approximately $0.30$, that it fluctuated over the three experimental sessions, and that the values were highest in the second experimental session (see Table \ref{Table2} in this Supplement). We followed this large-scale calculation with an investigation into the relationship between nodal flexibility (in particular brain regions) and learning. We found, as shown Figure 4 of the main manuscript, that the flexibility of a large number of brain regions could be used to predict learning in the following session. Here we also note that these regions were not those with highest flexibility or lowest flexibility in the brain. In fact, the flexibility of those regions that predicted learning was not significantly different from the flexibility of those regions that did not predict learning: $t \approx 0.01$ $p \approx 0.98$ (Session 1) and $t \approx 0.87$, $p \approx 0.38$ (Session 2).

In addition to those results reported in the main manuscript, we tested whether the flexibility of the cortical networks was significantly different from the flexibility expected in the (connectional, nodel, and temporal) random network null models. As we show in Table \ref{Table3} in this Supplement, the flexibility of the connectional and nodal null model networks was significantly higher than that of the cortical networks, and we found no discernible differences between the cortical networks and the temporal null model networks.  We found the greatest degree of flexibility in the nodal null model, in which individual nodes in any given time window were coupled to randomly selected nodes in the following time window. It is thus plausible that the subsequent disruption of nodal identity caused nodes to change computed module allegiances in this null model.


\paragraph{Robustness to Alternative Definitions}

It is important to assess the robustness of our findings to different definitions of flexibility. We therefore defined an alternative flexibility measure $f'_{i}$ to be the number of communities (modules) to which a node belongs at some point in a given experimental session. The mean alternative flexibility $F'$ is then given by averaging $f'_{i}$ over all nodes in the network: $F' = \frac{1}{N} \sum_{i=1}^{N} f'_{i}$. Using this alternative definition of flexibility, we again tested for differences between the cortical network and the three random network null models. As shown in Tables \ref{Table2} and \ref{Table3}, the $F'$ values of cortical networks were also significantly different from those in the null model networks. Interestingly, for this alternative definition, the temporal network null model exhibits significantly lower flexibility than the cortical networks, suggesting that this measure of flexibility might be sensitive to biologically relevant temporal evolution of modular architecture. Finally, we tested whether this alternative definition of flexibility also displayed a relationship to learning. Flexibility and learning were not significantly correlated in Session 1 ($r \approx 0.02$, $p \approx 0.90$) or in Session 2 ($r \approx 0.18$, $p \approx 0.48$), but flexibility in Session 1 was predictive of learning in Session 2 ($r \approx 0.64$, $p \approx 0.002$), and flexibility in Session 2 was predictive of learning in Session 3 ($r \approx 0.51$, $p \approx 0.019$).  These results for the alternative flexibility $F'$ are consistent with those of the original definition $F$, suggesting that our findings are robust.


\newpage
\newpage

\section*{Supplementary Discussion}
\addcontentsline{toc}{section}{Supplementary Discussion}


\subsection*{Resolution Limit of Modularity}
\addcontentsline{toc}{subsection}{Resolution Limit of Modularity}

When detecting communities by optimizing modularity and similar quality functions, it is important to note that modularity suffers from a resolution limit \cite{Fortunato2007,Fortunato2010,Porter2009,Good2010}.  As a result, the maximum-modularity partition can be biased towards a particular module size and can have difficulty resolving modules smaller than that size.  Consequently, small modules of potential interest have the potential to be hidden within larger groups of nodes that have been detected.  Modularity's resolution limit is particularly prevalent in sparse networks, binary networks, and large networks, and its effects tend to be much less significant in networks of the type (dense, weighted, and small) that we have studied \cite{Good2010}.


\subsection*{Measuring Differences in Brain States}
\addcontentsline{toc}{subsection}{Measuring Differences in Brain States}

In the present work, we have characterized differences in brain states during learning by examining the global network architecture and measuring differences between that architecture over three experimental sessions. An alternative line of investigation would be to seek network motifs (i.e., small patterns of nodes and edges) that have the potential to distinguish between brain states.  This could be done using statistical methods \cite{Zalesky2010b}, machine-learning techniques \cite{Richiardi2010}, or a combination of the two \cite{Shinkareva2010}. Our approach, however, has the advantage of assessing alterations in large-scale achitectural properties rather than differences in small parts of that architecture. Additionally, the approach that we have chosen provides a direct characterization of the underlying functional connectivity architecture irrespective of differences between brain states. Using this approach, we have therefore been able to demonstrate, for example, that there is significant non-random modular organization across multiple temporal scales.

\subsection*{A Note on Computational Time}
\addcontentsline{toc}{subsection}{A Note on Computation Time}

The investigations that we reported in the present work involved about $10,000$ CPU-days, and our study was therefore made possible by the use of two computing clusters available at the Institute for Collaborative Biotechnologies at UC Santa Barbara. Cluster 1 was composed of 42 Dell SC1425s (dual single-core Xeon 2.8GHz, 4GB memory), 5 Dell PE1950s (dual quad-core Xeon E5335 2.0GHz, 8GB memory), 1 Dell 2850 (RAID storage includes 500GB for the home directory), and MATLAB$^{\textregistered}$ MDCE with 128 worker licenses (cluster currently has 124 compute cores), Gigabit Ethernet, Software RAID backup node (converted compute node) with 673GB software RAID backup. Cluster 2 was composed of 20 HP Proliant DL160 G6s (dual quad-core E5540 ``Nehalem'' 2.53GHz, 24GB memory), 1 HP DL180 G6 (RAID storage includes 2.1TB for the home directory), MATLAB$^{\textregistered}$ MDCE with 160 worker licenses (cluster currently has 160 compute cores), Gigabit Ethernet, and a storage node with 4.6TB of RAID storage (for backup).

We performed maximization of the quality functions ($Q_{pm}$, $Q_{ml}$) a total of $100$ times for every connectivity matrix under study. In the static connectivity investigation, we constructed connectivity matrices for 20 subjects, 3 temporal scales (encompassing 1 experiment, 3 experimental sessions, and 25 time windows), and 1 random network null model. In the dynamic connectivity investigation, we constructed connectivity matrices for 20 subjects, 18--34 time windows, 3 different null models, 10 values of the inter-layer coupling $C$, and 4 values of time window length (80, 90, 100, and 110 TRs). In light of the computational extent of this work, we note that we did not employ Kernighan-Lin (KL) node-swapping steps \cite{Kernighan1970} in our optimization of $Q_{pm}$ or $Q_{ml}$, as they would be computationally prohibitive and are not necessary in the present context. KL steps move individual nodes between communities in order to further optimize a single sample of $Q_{pm}$ or $Q_{ml}$ \cite{Richardson2009,Newman2006PRE,Porter2009}. As we focus on the mean properties of  ensembles of partitions (and use them to report reliable measurements of architectural properties) rather than on the values of diagnostics for any individual partitions, KL steps that provide a marginal increase in the value of $Q_{pm}$ or $Q_{ml}$ would not be helpful for our study.

\newpage
\newpage
\newpage
\small
\begin{table}
\begin{center}
\begin{tabular}{| l | l |}
\hline
Frontal pole &	Cingulate gyrus, anterior \\
Insular cortex&	Cingulate gyrus, posterior \\
Superior frontal gyrus & Precuneus cortex \\
Middle frontal gyrus & Cuneus cortex \\
Inferior frontal gyrus, pars triangularis & Orbital frontal cortex \\
Inferior frontal gyrus, pars opercularis &Parahippocampal gyrus, anterior \\
Precentral gyrus &Parahippocampal gyrus, posterior \\
Temporal pole & Lingual gyrus \\
Superior temporal gyrus, anterior &Temporal fusiform cortex, anterior \\
Superior temporal gyrus, posterior& Temporal fusiform cortex, posterior \\
Middle temporal gyrus, anterior &Temporal occipital fusiform cortex \\
Middle temporal gyrus, posterior &Occipital fusiform gyrus \\
Middle temporal gyrus, temporooccipital& Frontal operculum cortex \\
Inferior temporal gyrus, anterior & Central opercular cortex \\
Inferior temporal gyrus, posterior & Parietal operculum cortex \\
Inferior temporal gyrus, temporooccipital 	&Planum polare \\
Postcentral gyrus	&Heschl's gyrus \\
Superior parietal lobule	&Planum temporale \\
Supramarginal gyrus, anterior&	Supercalcarine cortex \\
Supramarginal gyrus, posterior	&Occipital pole \\
Angular gyrus	&Caudate \\
Lateral occipital cortex, superior	&Putamen \\
Lateral occipital cortex, inferior	&Globus pallidus \\
Intracalcarine cortex	&Thalamus \\
Frontal medial cortex	&Nucleus Accumbens \\
Supplemental motor area&	Parahippocampal gyrus (superior to ROIs 34,35) \\
Subcallosal cortex	&Hippocampus \\
Paracingulate gyrus	&Brainstem \\
\hline
\end{tabular}
\end{center}
\caption{\footnotesize \textbf{Brain regions present in the Harvard-Oxford Cortical and Subcortical Parcellation Scheme provided by FSL \cite{Smith2004,Woolrich2009}.}  \label{Table1}}
\end{table}
\normalsize

\newpage
\small
\begin{table}
\begin{center}
\begin{tabular}{| l | l | l | l | l | l |}
\hline
Type of Flexibility & Session 	& Cortical		& Connectional 	& Nodal 	& Temporal \\
\hline
$F$ & 		& ~ 			& ~ 		& ~ 		& ~	\\
\hline
~ & 1		& 0.027$\pm$0.009	& 0.041$\pm$0.008 & 0.070$\pm$0.001 & 0.027$\pm$0.010 \\
~ & 2		& 0.030$\pm$0.009	& 0.042$\pm$0.007 & 0.067$\pm$0.001 & 0.030$\pm$0.009 \\
~ & 3		& 0.027$\pm$0.008	& 0.039$\pm$0.008 & 0.064$\pm$0.001 & 0.026$\pm$0.009 \\
\hline
$F'$ &		& ~ 		& ~ 		& ~ 		& ~	\\
\hline
~ & 1	 	& 1.93$\pm$0.30		& 2.58$\pm$0.25	& 3.19$\pm$0.01 & 1.90$\pm$0.31 \\
~ & 2	 	& 1.95$\pm$0.27		& 2.53$\pm$0.23	& 3.01$\pm$0.01	& 1.94$\pm$0.28 \\
~ & 3	 	& 1.86$\pm$0.25 	& 2.43$\pm$0.23	& 2.92$\pm$0.01	& 1.85$\pm$0.26 \\
\hline
\end{tabular}
\end{center}
\caption{\footnotesize \textbf{Flexibility in cortical network and random network null models} for two different definitions of flexibility. One definition is $F = \frac{1}{N} \sum_{i=1}^{N} f_{i}$, where $f_{i}$ is the number of times node $i$ changes module allegiance in a given experimental session divided by the total number of possible changes (i.e., by the number of consecutive pairs of layers in the multilayer framework). The second definition is $F' = \frac{1}{N} \sum_{i=1}^{N} f'_{i}$, where $f'_{i}$ is defined as the total number of modules of which node $i$ is a member at some point in the experiment session.  We also indicate mean values and standard deviations. \label{Table2}}
\end{table}
\normalsize

\newpage
\small
\begin{table}
\begin{center}
\begin{tabular}{| l | l | l | l | l |}
\hline
Type of Flexibility 	& Session 	& Null Model  	& $t$-statistic 	& $p$-value \\
\hline
$F$ 			& 1 		& Connectional	& $-10.62$	& $1.2 \times 10^{-18}$ \\
~	 		& ~ 		& Nodal		& $-47.56$	& $1.1 \times 10^{-75}$ \\
~	 		& ~ 		& Temporal	& $-0.64$		& 0.51 \\
\hline
~	 		& 2 		& Connectional	& $-9.47$		& $5.5 \times 10^{-16}$ \\
~	 		& ~ 		& Nodal		& $-43.41$	& $1.7 \times 10^{-71}$ \\
~	 		& ~ 		& Temporal	& $-0.16$		& 0.86 \\
\hline
~	 		& 3 		& Connectional	& $-9.86$		& $1.1 \times 10^{-16}$ \\
~	 		& ~ 		& Nodal		& $-46.95$	& $4.5 \times 10^{-75}$ \\
~	 		& ~ 		& Temporal	& 0.77		& 0.43 \\
\hline
\hline
$F'$	 		& 1 		& Connectional	& $-14.54$	& $1.8 \times 10^{-27}$ \\
~	 		& ~ 		& Nodal		& $-43.44$	& $1.6 \times 10^{-71}$ \\
~	 		& ~ 		& Temporal	& 4.67		& $8.4 \times 10^{-6}$ \\
\hline
~	 		& 2 		& Connectional	& $-14.04$	& $2.3 \times 10^{-26}$ \\
~	 		& ~ 		& Nodal		& $-39.67$	& $2.1 \times 10^{-67}$ \\
~	 		& ~ 		& Temporal	& 3.95		& $1.3 \times 10^{-4}$ \\
\hline
~	 		& 3 		& Connectional	& $-14.87$	& $3.5 \times 10^{-28}$ \\
~	 		& ~ 		& Nodal		& $-43.80$	& $6.9 \times 10^{-72}$ \\
~	 		& ~ 		& Temporal	& 2.79		& 0.0061 \\
\hline
\end{tabular}
\end{center}
\caption{\footnotesize \textbf{Results for one-sample $t$-tests on flexibility in cortical networks versus random network null models} for two different definitions of flexibility ($F$ and $F'$).  We report these results for the connectional, nodal, and temporal null models. Negative $t$-statistics indicate that the flexibility of the null model network is greater than that of the cortical network, and positive $t$-statistics indicate that the flexibility of the cortical network is greater than that of the null model network. \label{Table3}}
\end{table}
\normalsize


\newpage
\newpage
\newpage

\newpage
\newpage
\newpage

\newpage
\newpage
\newpage

\newpage
\newpage
\newpage



\newpage
\newpage
\newpage
\bibliography{bibfile}


\section*{Addendum}

This work was supported by the David and Lucile Packard Foundation, PHS Grant NS44393, the Institute for Collaborative Biotechnologies through contract no. W911NF-09-D-0001 from the U.S. Army Research Office, and the NSF (DMS-0645369).  M.A.P. acknowledges a research award (\#220020177) from the James S. McDonnell Foundation.  \\
We thank Aaron Clauset for useful discussions and John Bushnell for technical support.\\
Competing Interests: The authors declare that they have no competing financial interests. \\
Correspondence: Correspondence and requests for materials should be addressed to D.S.B. (email: dbassett@physics.ucsb.edu).\\
